# Neural Memory Orchestration

End-to-End Learned Page Prefetching with Disaggregated Inference

Michael Zizov
Afeka Academic College of Engineering
in Tel Aviv, Israel
mizizov@gmail.com

Boris Morose
Afeka Academic College of Engineering
in Tel Aviv, Israel
borism@s.afeka.ac.il

## ABSTRACT

Modern HPC workloads increasingly stall on the memory hierarchy rather than on compute, and page faults from mmap'd files, swap, and tiered memory are a growing fraction of wall time. Heuristic prefetchers in the kernel handle stride-like and temporally clustered patterns well but degrade sharply on the irregular, data-dependent access streams typical of these workloads. Learned prefetchers have been shown to capture such patterns in simulation but have not been taken to a deployed pipeline running against live workloads on real hardware. We present Overseer, an end-to-end learned page-prefetching system. An eBPF collector observes page faults on the target host, a transformer predictor (**OverNet**) runs on network-attached inference hardware and emits future page addresses, and a kernel-resident enforcer materializes those pages into the target's address space before they are demanded. **OverNet** encodes fault sequences through per-instance ordinal ranks over deltas, which lets a single trained model serve any target process without retraining. On two page-fault-heavy workloads, Overseer reduces runtime by roughly 35% on Hilbert-curve traversal and by roughly an order of magnitude on dense matrix multiplication, with corresponding reductions in major-fault counts.

## CCS CONCEPTS

Computer systems organization → Memory architectures;

## KEYWORDS

page prefetching, learned prefetching, memory access prediction, memory systems, irregular memory access, high-performance computing, eBPF, transformer models, Linux kernel

## 1 Introduction

Modern HPC workloads are increasingly bottlenecked by the memory hierarchy. Large-scale ML training is one of the most visible examples, its CPU-resident phases generating dense, irregular fault streams; scientific simulation shows the same pattern [1]. As working sets outgrow physical RAM, each fault becomes a synchronous stall in which the running thread waits microseconds-to-milliseconds for a backing store.

Operating systems ship heuristic mechanisms for this regime, including readahead, NUMA migration, and MGLRU, but these target stride-like or temporally clustered patterns and degrade sharply on the irregular, data-dependent access streams characteristic of modern HPC applications. Learned prefetchers [2, 3, 4] capture these patterns where heuristics fail, and recent work [5] has narrowed the gap to practical latency budgets. Two limitations have kept this line of work from production. First, it has been studied almost exclusively in trace-driven simulation, deferring the systems questions of how the predictor reaches the kernel and what happens when inference cannot keep up with the access stream. Second, it targets cache lines rather than pages. Cache misses are nanoseconds and are absorbed by hardware; page faults are microseconds-to-milliseconds and are absorbed by the running thread, making page-level predictions both an underexplored target and a more natural fit for the expensive sequence models ML-for-systems work has been producing.

We present **Overseer**, an end-to-end learned page-prefetching system. To our knowledge, Overseer is the first learned memory prefetcher taken all the way to a deployed pipeline running against live workloads on real hardware. Page faults are observed live on the host by the **Collector**, sequences are streamed to **OverNet** on attached external hardware for inference, and predicted pages are materialized into the target's address space by a kernel-resident **Enforcer** before they are demanded. The disaggregated design isolates inference latency from the host's compute path, lets the model run on hardware co-designed for it, and keeps kernel-resident components small.

Overseer makes the following contributions:

1. **An end-to-end learned page-prefetching system.** A complete pipeline covering observation, off-host neural inference, and kernel-resident enforcement, running on real hardware against real workloads, where prior work has been almost exclusively simulation-only.

2. **OverNet**, a neural network trained for page-fault prediction, adapted to a real-time deployment budget on dedicated hardware.

3. **A characterization of disaggregated inference**: the latency and throughput conditions under which off-host prediction is viable.

## 2 Related Work

**Heuristic prefetchers.** Decades of work on rule-based prefetchers, including stride, Best-Offset [6], ISB [7], and Domino [8], have produced strong baselines for regular and temporally correlated access patterns and continue to be deployed in production CPUs and operating systems. However, they falter on the irregular, data-dependent patterns characteristic of modern HPC and ML workloads, which is precisely the gap that learned prefetchers aim to address [2, 3, 4, 5].

**Neural cache-line prefetching.** Hashemi et al. [2] first cast prefetching as sequence prediction using an LSTM, and Voyager [3] extended this approach with a hierarchical architecture that separates page and offset prediction. Subsequent work has explored attention-based designs (TransFetch [4]) and lightweight, customized models that combine clustering with history tables (Twilight and T-LITE [5]). These approaches target the cache line as the unit of prediction and report results using simulators such as ChampSim [3, 4, 5], rather than running as end-to-end systems on real hardware. **Overseer** differs in both respects: we predict at page granularity, where the latency budget permits a substantially larger model, and we evaluate an end-to-end deployed pipeline rather than relying on trace replay.

**Learned page-level prefetching.** Page-level neural prefetching remains comparatively underexplored. The closest work is Long et al. [9], which applies a Transformer to page prefetching in CPU-GPU Unified Virtual Memory (UVM). That work targets a different memory regime, UVM, with driver-level involvement, and remains simulation-based. In contrast, Overseer operates on host-RAM page faults serviced by the Linux kernel and runs end-to-end on real hardware.

**Dedicated hardware for learned memory policies.** LSTM-CRP [10] implements a learned cache-replacement policy on a custom on-die FPGA accelerator. Overseer shares the idea of offloading inference to specialized hardware but takes the opposite architectural choice: its inference engine is network-attached rather than integrated on-die, which enables a substantially larger model than an on-die accelerator could accommodate.

**Transformer architecture.** The Transformer architecture, introduced by *Attention Is All You Need* [11], forms the basis of the model used by the **OverNet** inference engine. The Transformer architecture provides mechanisms such as sinusoidal positional encodings, which **OverNet** uses to represent the position of each page within the input sequence before processing the sequence through its encoder.

## 3 Methodology

### 3.1 Design Overview

**Overseer** is a three-stage pipeline that observes page faults, predicts future page accesses, and prefetches the corresponding pages for a designated target process (or set of processes). The three stages are physically and conceptually decoupled.

The **Collector** runs on the same host as the target process. It observes userspace page faults, filters those belonging to the target's Cgroup, and assembles per-PID sliding windows of recent fault events. When a window fills, it is sent over UDP to the inference appliance.

**OverNet** runs on network-attached dedicated hardware. It receives a fault window, encodes it into ordinal ranks over the window's IPs and pages, runs inference, and emits a sequence of predicted future page addresses. The predictions, together with the originating PID, are returned to the host over UDP.

The **Enforcer** is a kernel module on the target's host. It receives **OverNet**'s output packets and dispatches each to a per-PID kernel worker attached to the corresponding process's mm; the worker forces each predicted page into residency.

The components are coupled by a deliberately narrow contract. The **Collector** emits (PID, IP_window, page_window) tuples; predictions returned to the **Enforcer** are (PID, predicted_page_sequence) tuples. Nothing in the contract is workload or model specific: the same **Collector** and **Enforcer** can serve any predictor that respects the contract, and the same predictor can serve any host that produces compatible fault windows. The data path is fixed and workload-agnostic; the predictor's learned weights are the only workload-specialized element.

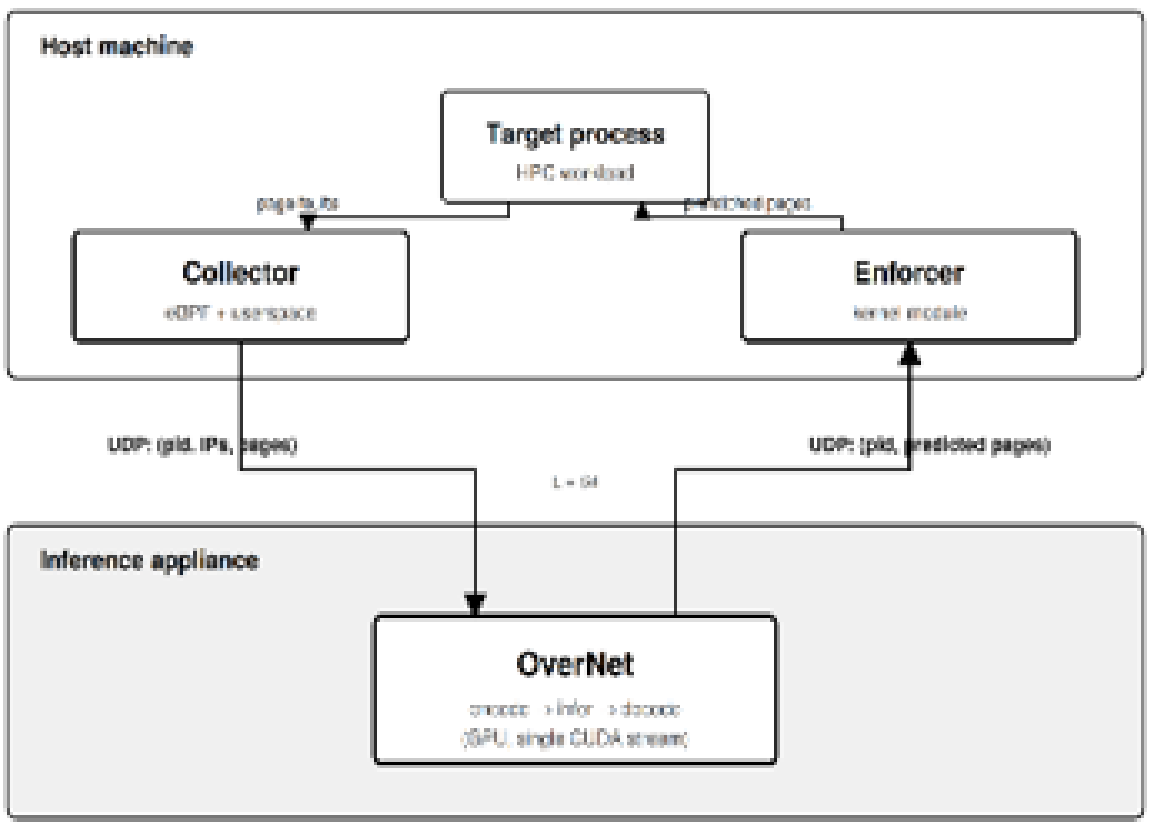

Figure 1: **Overseer system overview.**

## 3.2 The Collector

The **Collector** has a kernel-space component and a user-space component. An eBPF component intercepts every user-space page fault and forwards qualifying events, comprising the faulting instruction pointer (IP) and the effective page address, to user space. The user-space component aggregates these events into per-PID sliding windows of length L. When a window fills, the Collector emits a packet containing the PID, the L IPs, and the L page addresses, and starts collecting the next window for that PID.

Per-PID windowing is required because the model expects sequence inputs reflecting a single execution context; interleaving faults from multiple processes would corrupt the input distribution. A multi-process target, typical of MPI-style HPC jobs, therefore produces one independent window stream per PID from a single Collector instance.

Batching amortizes both the per-window network round-trip, and the GPU launch overhead at the inference appliance, while giving the model enough context to capture patterns rather than merely nearest-neighbor correlations.

**Design tradeoffs.** The window length L trades context length against responsiveness: longer windows give the predictor more history and more output coverage per inference but increase the delay between a fault occurring and the corresponding prediction being requested. We chose L=64 as a compromise informed by both training-time accuracy curves and observed fault inter-arrival times; we revisit this choice in Section 5. The Collector maintains no prediction state, so a Collector failure degrades gracefully: no predictions arrive, but the target's correctness remains unaffected.

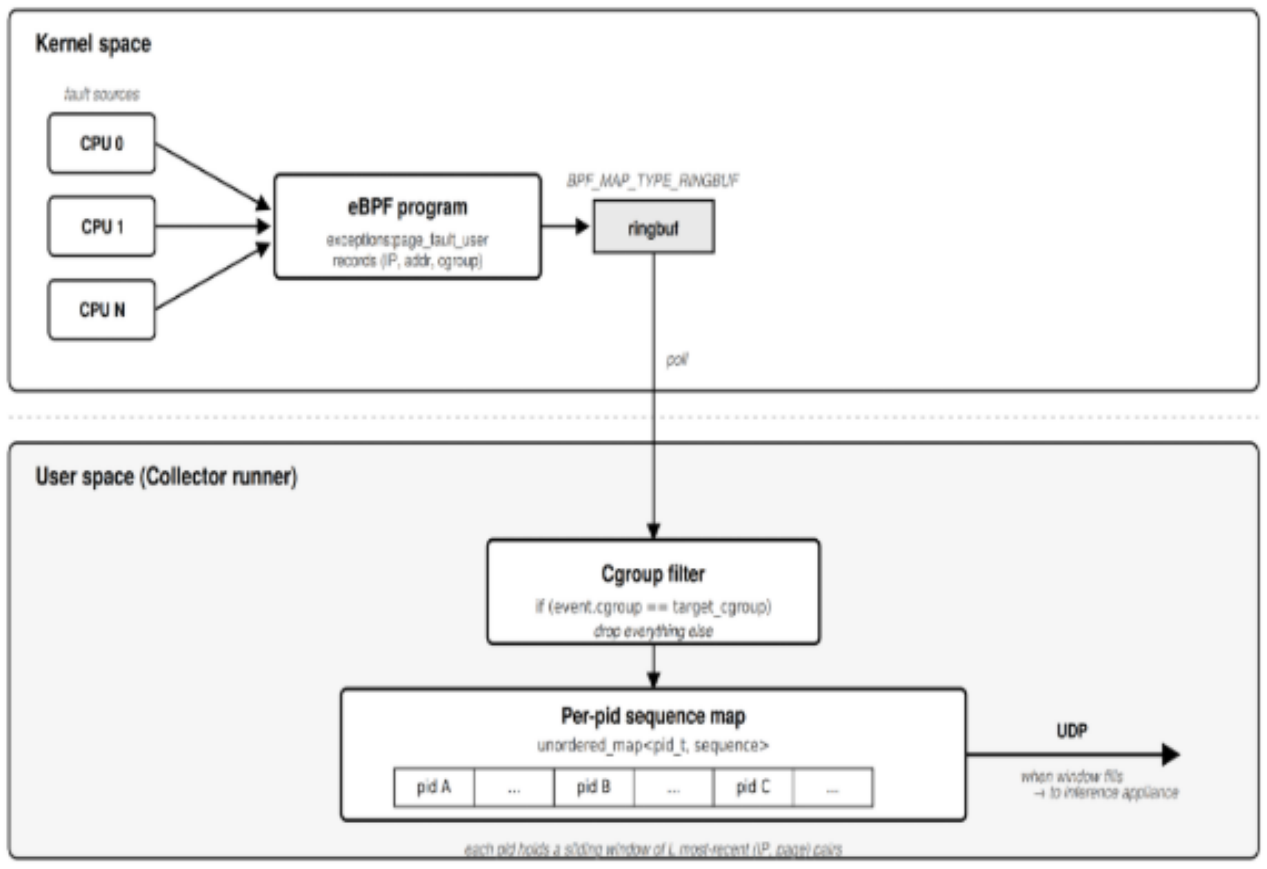

Figure 2: **The Collector architecture.**

## 3.3 OverNet

**OverNet** is the deep-learning-based predictor at the center of the pipeline**.** It takes a fault window, a sequence of $L$ recent (IP, page) pairs from a single process and emits a sequence of $L$ predicted future pages.

**Input encoding: deltas mapped to online ordinal ranks.** One of **Overseer's** most consequential representational choices is how addresses are presented to the model. Prior work has used absolute embeddings of pages and offsets [3], deltas mapped to a fixed offline vocabulary [2], or bitmap outputs over a bounded delta range [4]. Each of these encodings ties the model either to a specific training trace's address layout or to a specific spatial-distance window. **OverNet** uses a different approach: it operates in delta space but constructs its delta vocabulary online and on a per-instance basis.

The encoding proceeds in two stages. First, each input stream, the IP stream and the page stream is converted into a delta sequence. Given a page sequence

$$p_0, p_1, \cdots, p_{L-1}$$

The delta sequence is defined as $\Delta_0 = 0$ as a sentinel anchor, followed by

$$\Delta_i = p_i - p_{i-1} \quad i \geq 1$$

The same construction is applied to the IP sequence. Second, each delta is mapped to an integer rank using a per-instance LookupTable: the $k$-th distinct delta to appear in the target's history is assigned rank $k$. Subsequent occurrences of the same delta retrieve the same rank. The LookupTable persists for the lifetime of the target instance and is shared across all its fault windows; IPs and pages have separate tables. The model's input tokens are therefore ranks over the delta vocabularies built up so far for this specific target.

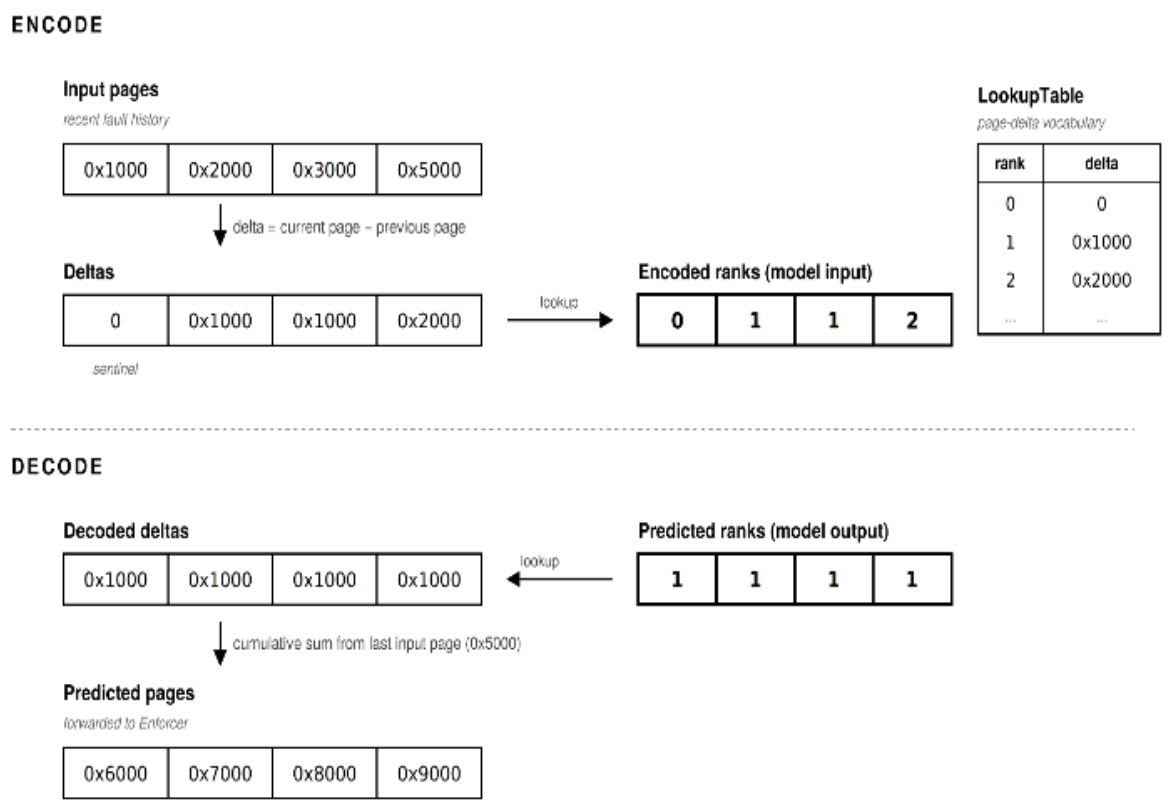

Figure 3: **Encoding and decoding example for L=4.**

Encoding deltas rather than raw addresses makes online vocabulary construction practical. The set of distinct addresses touched by a workload grows roughly linearly with execution time; a strided scan over a large array produces thousands of unique pages even though the access pattern is structurally trivial. The set of distinct deltas between consecutive accesses is substantially smaller and converges quickly: a strided scan of any length contributes only a single delta. In our deployment, the LookupTable's growth rate drops sharply within the first few thousand faults of a typical workload.

Assigning ranks online rather than offline makes the model deployable across processes without retraining. The model never sees an absolute delta value, only its rank in a vocabulary constructed by the target instance. The semantic meaning of "rank 3" is consistent across instances: it represents the third distinct jump distance observed for that target, even if the underlying delta value differs entirely. This ordinal semantic is what the model learns to operate on, and what allows a single trained OverNet to serve different target instances.

**Model architecture. OverNet** is a Transformer encoder-decoder that consumes IP-rank and page-rank tokens as two parallel input sequences and emits a sequence of output page-rank tokens (Figure 4). The IP stream serves as conditioning, providing the code context associated with each fault; repeated faults from the same call site can signal a structured access pattern. Only the page-rank sequence is emitted.

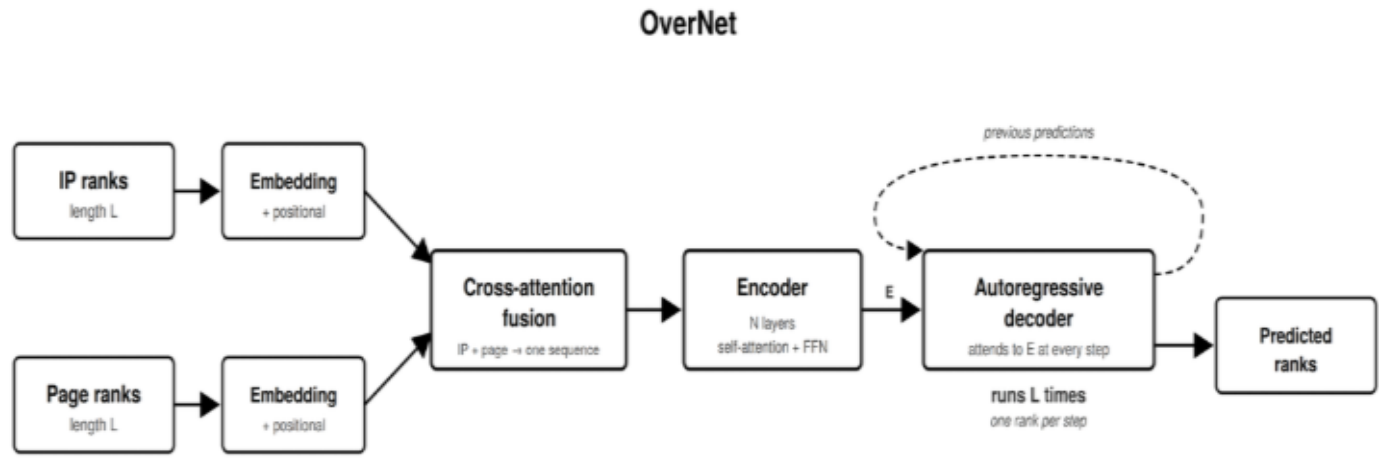

Figure 4: **The OverNet neural network architecture.**

**Inference pipeline.** All OverNet operations, including encoding, inference, and decoding, run on the GPU within a single CUDA stream. The decode step takes the predicted rank sequence, maps each rank back to its delta via the page LookupTable, and cumulatively sums the deltas from the last observed absolute page to reconstruct the predicted absolute page addresses.

**Design tradeoffs.** The dominant tradeoff is between model capacity and inference latency. Larger models predict more accurately but increase per-window inference time, delaying the prediction's arrival at the **Enforcer** and reducing the window over which the prediction remains useful. The disaggregated architecture substantially relaxes this tradeoff compared with on-die designs: we have roughly three orders of magnitude more latency budget than a cache-line prefetcher, though the tradeoff is not eliminated. A secondary tradeoff concerns the vocabulary cap V: too low forces frequent eviction of older deltas, too high enlarges the SoftMax output space and increases inference latency. When the table is full, the oldest entry (by insertion order) is overwritten. A separate limitation is that the model cannot predict deltas absent from the LookupTable; this constrains the startup phase of any new target instance, and we characterize it in Section 5.

## 3.4 The Enforcer

**The Enforcer** is the kernel-resident component that turns **OverNet's** predictions into actual page residency in the target's address space.

A predicted page is not yet a prefetched page. For a prediction to provide any benefit, the kernel must traverse the prediction list and ensure that each page is resident in the target process's memory before the target faults on it. The **Enforcer** maintains a pool of per-PID kernel worker threads. When a prediction packet arrives:

- The **Enforcer** dispatcher looks up the PID in its worker table.
- If no worker exists for that PID, a new worker is spawned and attached to the target's mm (memory descriptor).
- The **Enforcer** places the newly received prediction sequence into the corresponding worker thread's queue.
- Each thread continuously monitors its queue for new prediction sequences and processes each predicted page in sequence.
- The worker traverses the predicted page list and forces each page into residency.

The worker remains alive after processing the prediction to amortize the cost of mm attachment across subsequent prediction packets for the same PID.

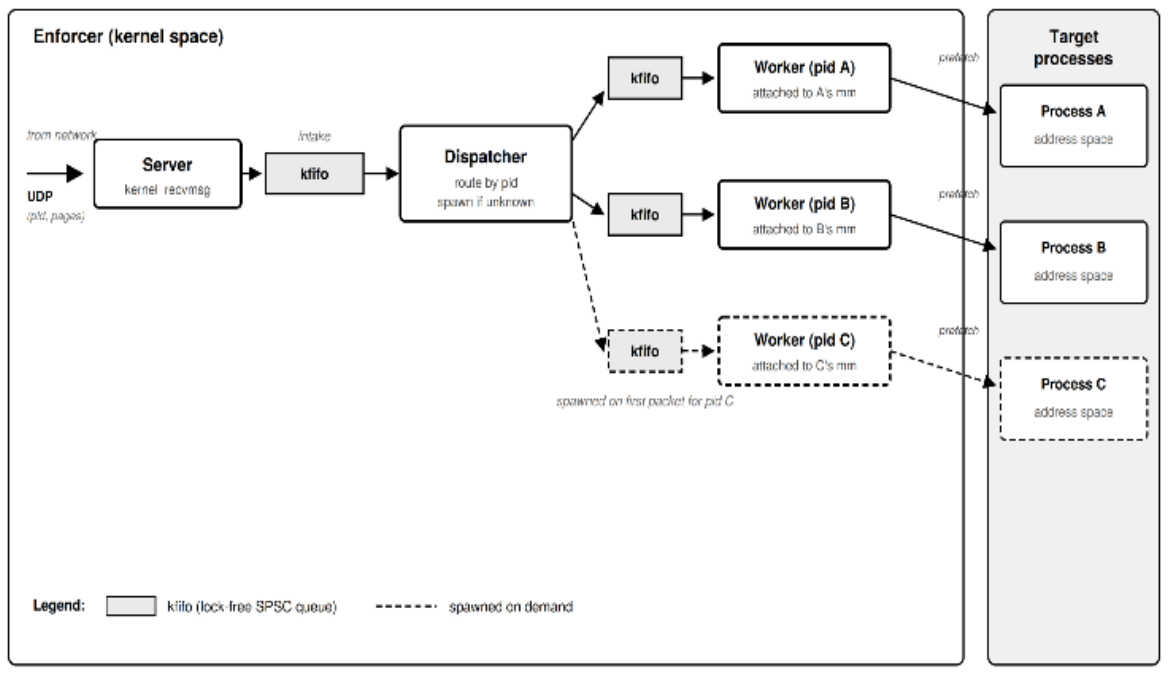

Figure 5**: Enforcer module architecture**

Per-PID workers are necessary because forcing page residency operates on a specific memory descriptor (mm), and each worker must be attached to the correct address space to perform prefetching. Spawning workers on demand, rather than maintaining a fixed pool, allows the **Enforcer** to adapt to the number of processes exposed by the target without requiring static configuration.

**Design tradeoffs.** The prediction channel is UDP rather than Netlink. The latter would provide a clean in-kernel delivery path with no network trust surface, but it is confined to a single host; choosing it would force OverNet on-die with the target and forfeit the disaggregation that keeps inference latency off the host's compute path. UDP preserves disaggregation at the cost of a trust boundary between hosts; authenticated framing on that link is left to future work.

## 4. Implementation

This section describes the concrete implementation of each component. Whereas Section 3 explains the design decisions, Section 4 describes the technologies, kernel interfaces, and engineering choices used to realize the running system.

**Hardware and software stack.** The host runs Linux and executes the target workload, the **Collector**, and the **Enforcer**. The dedicated inference platform is an NVIDIA Jetson Orin Nano running JetPack and TensorRT. The host and the inference platform are connected via Ethernet and communicate using UDP.

## 4.1 Collector

The **Collector** is implemented in C++17 using a libbpf-generated BPF skeleton. The kernel-space component is an eBPF program attached to the exceptions:page_fault_user tracepoint, which fires on every user-space page fault on every CPU. For each event, the program records three fields: the faulting instruction pointer, the effective address, and the Cgroup ID of the faulting task. It then writes a fixed-size record to a BPF_MAP_TYPE_RINGBUF shared with user space. The program is deliberately minimal; read, pack, and write, to keep its verifier footprint small and its per-event overhead bounded.

The user-space component runs as a single C++ process. The **Collector** is invoked with the target's Cgroup ID as a command-line argument and stores it as a filter constant. It then polls the ring buffer in a loop. For each event, it compares the event's Cgroup ID with the configured target ID; non-matching events are dropped, while matching events are dispatched to the corresponding per-PID state. Filtering in user space rather than in the eBPF program simplifies the program by avoiding additional BPF maps and helper logic, at the cost of forwarding non-target events to user space. In practice, this overhead is acceptable because the filtering operation requires only a single integer comparison and each eBPF event is small.

Per-PID state is maintained in an std::unordered_map keyed by PID. Each entry holds a fixed-length sequence of L recent (IP, page) pairs (L = 64 in our implementation). When a PID's sequence fills, its contents are serialized into a UDP datagram destined for the inference hardware, after which the sequence is reset to collect the next L faults.

## 4.2 OverNet

**OverNet** is trained in PyTorch and exported to TensorRT via ONNX for deployment on the inference hardware. The deployed inference pipeline is a single CUDA stream comprising the encode, inference, and decode kernels described in §3.3. The two LookupTables (one for IPs, one for pages) for each target pid are allocated in GPU global memory at the size of the vocabulary cap $V$ and persist across all forward passes for that target. They are keyed in a host-side map by pid; a new packet's pid selects the corresponding pair of tables for that invocation of the stream.

The LookupTable is a closed-form hash table mapping deltas to ranks, paired with a rank-indexed buffer of the current vocabulary contents. The encode kernel launches a single block of $L$ threads, one per token in the window that cooperatively performs delta computation and rank lookup. Each thread computes its delta, then looks up its delta in the table; threads whose deltas are already present read their rank in parallel and the kernel is essentially done. Threads whose deltas are absent leave a sentinel in the rank array. After a syncthreads barrier, a single designated thread walks the array and inserts any missing deltas sequentially, advancing a circular head pointer; when the head wraps past $V$, the oldest entry is overwritten and its corresponding hash slot tombstoned. This design is deliberately unsynchronized across threads: it relies on the empirical observation that in steady state most deltas in a window are already in the vocabulary, so the parallel-lookup phase covers the vast majority of the work and the sequential-insertion phase has little to do. Only during workload warmup, or on rare windows containing previously-unseen jump patterns, does the sequential phase materially affect kernel time.

The decode kernel runs an inclusive prefix sum over the predicted deltas and adds $p_{L-1}$ to produce the absolute predicted pages. Folding encode and decode into CUDA kernels  rather than performing them on the inference hardware's CPU avoids two PCIe-equivalent crossings per prediction and keeps the LookupTables resident on the GPU where the model needs them.

**OverNet** runs at fp16 precision. The exported TensorRT engine is built once per (model, $L$, $V$) configuration and loaded at startup; no rebuilding occurs at runtime.

## 4.3 Enforcer

The **Enforcer** is a loadable kernel module written in C using the Linux kernel API. It consists of three classes of kernel threads, all created via kthread_run and connected by kfifo ring buffers:

1. **Server thread** binds a UDP socket to a configured port and reads **OverNet** prediction packets directly in kernel space using kernel_recvmsg. Each packet containing a pid and a sequence of $L$ predicted page addresses is pushed into the *intake kfifo*.

2. **Dispatcher thread** (the "main" **Enforcer** thread) reads from the intake kfifo, looks up the originating pid in its worker table, and pushes the prediction into the corresponding *per-pid kfifo*. If no worker exists for this PID  the target has spawned a new process **Overseer** has not yet seen, the dispatcher spawns one via kthread_run, attaches it to the target's mm via get_task_mm, and creates its dedicated kfifo before forwarding the prediction.

3. **Per-PID worker threads** each loop on their own kfifo, draining one prediction at a time. For each predicted page, the worker forces the page into residency in its attached mm. Workers persist across predictions to amortize the cost of mm attachment, which is non-trivial relative to the per-page work.

All inter-thread communication is via kfifo. The kernel's kfifo primitive provides lock-free single-producer/single-consumer queuing, which matches the **Enforcer's** communication graph exactly: each kfifo has exactly one writer thread and one reader thread, so no additional synchronization is required. Back pressure is implicit, a full kfifo causes the writer to drop the prediction rather than block, which is the safer failure mode for a kernel-resident component.

## 5. Results

We evaluate **Overseer** on two page-fault-intensive workloads, characterizing both its end-to-end impact on runtime and its effect on the underlying major-fault rate. Our evaluation addresses two central questions: (i) whether **Overseer** meaningfully reduces runtime on workloads dominated by demand paging, and (ii) how sensitive This reduction is to the latency of the inference path, which the disaggregated design puts on a network link rather than on the host CPU.

### 5.1 Experimental Setup

**Workloads.** We use two benchmarks, each configured so that its working set exceeds host RAM and is backed by an mmap'd file on disk. This forces the workload into a regime dominated by major page faults, which is precisely the regime **Overseer** is designed for.

- **Hilbert curve traversal.** A traversal of a large 2D array along a Hilbert space-filling curve. The access pattern is spatially local at short range but produces irregular, data-dependent jumps at longer range, which is difficult for stride-based readahead.
- **Dense matrix multiplication.** A matmul over mmap'd operands. Access is structured but the working set per block far exceeds physical RAM, so each block transition triggers a burst of major faults.

**Configurations.** We compare three configurations:

- **Baseline.** The workload runs on the host using the default Linux page cache and readahead. No Overseer components are loaded.
- **Overseer Local.** The full **Overseer** pipeline is active, with **OverNet** running on an NVIDIA RTX 4060 Ti co-resident with the target on the host. This configuration represents an upper bound on what **Overseer** can deliver when inference latency is minimized.
- **Overseer Remote.** The full **Overseer** pipeline is active, with **OverNet** running on a network-attached Jetson Orin Nano connected to the host via Ethernet. This is the deployment configuration described in §3 and §4.

Both configurations run the same **OverNet** weights, so any performance gap between them is attributable to the round-trip time and inference throughput of the remote path.

**Methodology.** For each (workload, configuration) pair we ran 30 independent trials, recording wall-clock runtime (elapsed_s) and the target's major-fault count (majflt) from /usr/bin/time -v. Runs were performed on an otherwise idle host with page cache flushed between trials to control cross-run caching.

### 5.2 Hilbert Curve

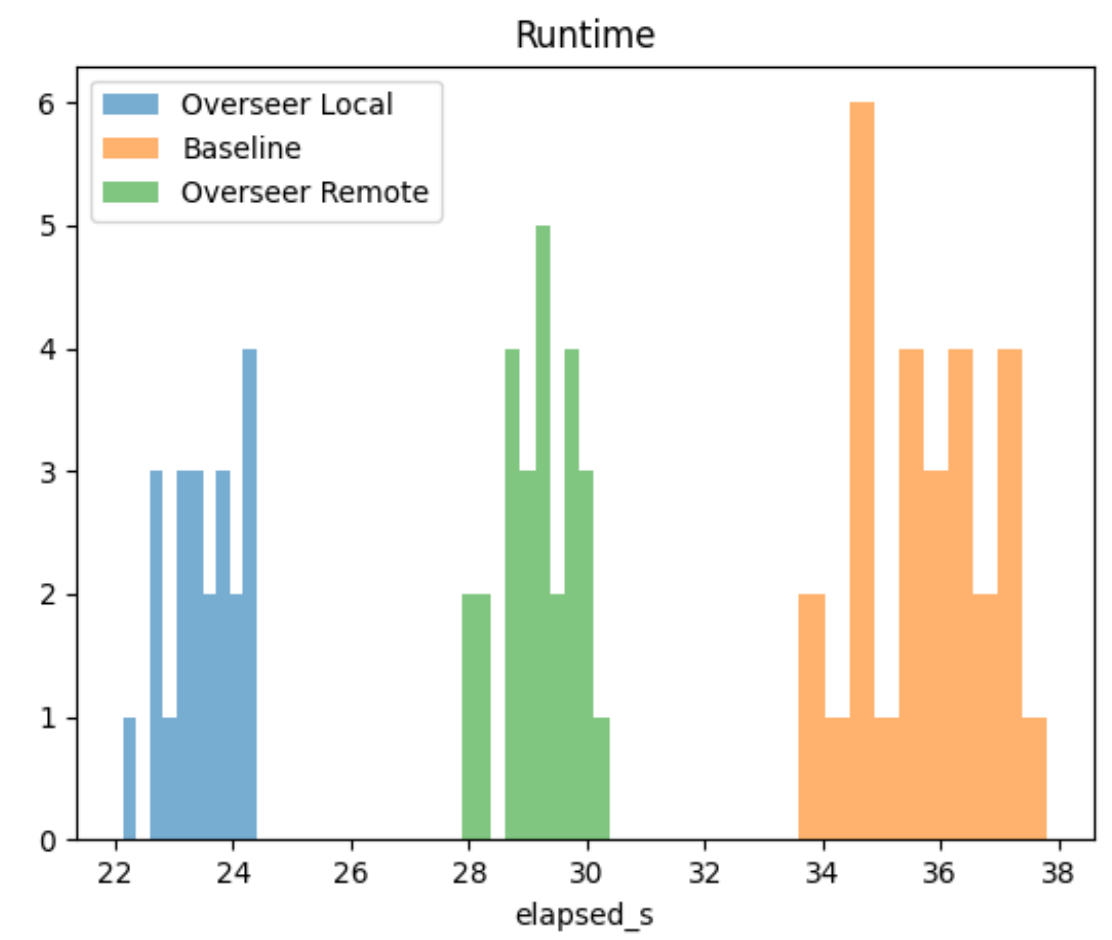

Figure 6: **Distribution of wall-clock runtimes for the Baseline, Overseer Local, and Overseer Remote configurations on the Hilbert-curve workload.**

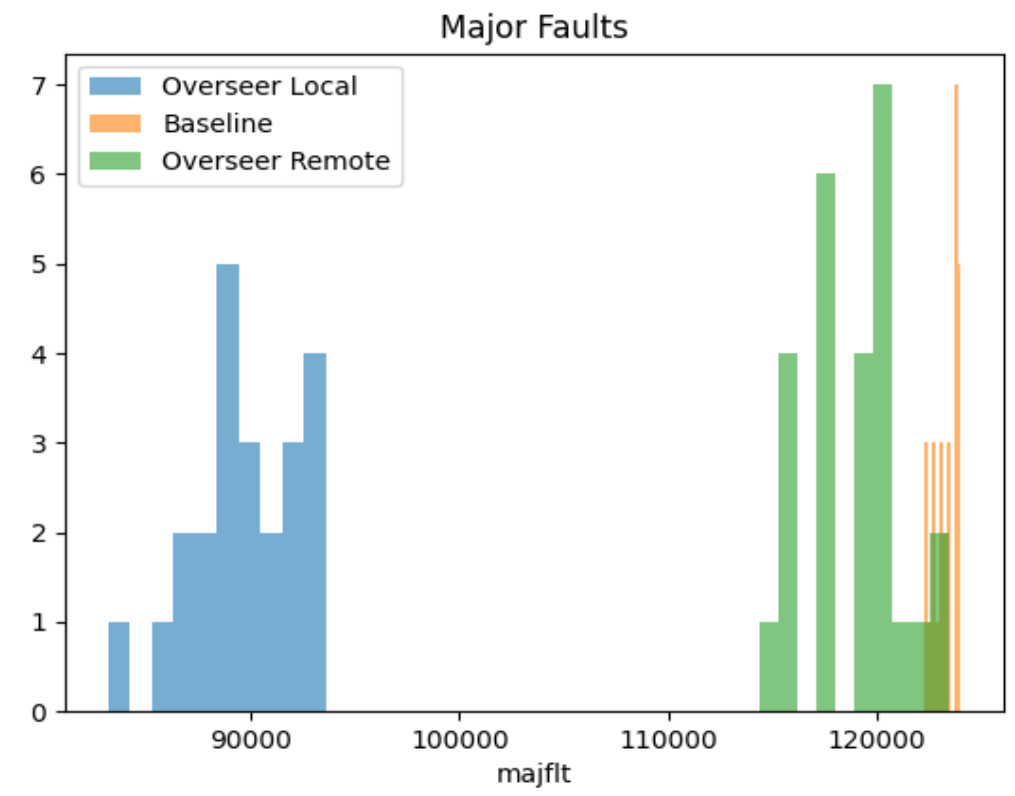

Figure 7: **Histogram of major page-fault counts for the Baseline, Overseer Local, and Overseer Remote configurations on the Hilbert-curve workload.**

Figures 6 and 7 show the runtime and major-fault distributions across the 30 trials for each configuration.

**Overseer Local** reduces mean runtime from approximately 35.5 s in the **Baseline** to approximately 23 s, corresponding to a reduction of roughly 35%. **Overseer Remote** reduces mean runtime to approximately 29s, a reduction of roughly 18%. The three distributions are clearly separated: no **Overseer Local** trial exceeds any **Baseline** trial, and only a small tail of **Overseer Remote** overlaps the fast end of the **Baseline** distribution.

The major-fault picture is more nuanced. **Overseer Local** reduces major faults from approximately 123k in the **Baseline** to approximately 90k, a reduction of roughly 27%: the predictor is anticipating a substantial fraction of the workload's faults and converting them into background-serviced page fills that no longer register as major faults on the target's thread. **Overseer Remote** reduces major faults only modestly, to approximately 118k. The disparity between the runtime improvement of the Remote configuration (18%) and its reduction in major faults (a few percent) warrants further examination. One possible explanation is that even when a prediction arrives too late to eliminate a fault entirely, it may still reduce the fault's service time by initiating retrieval of the backing page before the demand fault occurs. Thus, **Overseer's** benefit may arise not only from avoiding major faults but also from reducing their service latency. The Local configuration appears to benefit from both effects, whereas the Remote configuration appears to benefit primarily from the latter.

## 5.3 Dense Matrix Multiplication

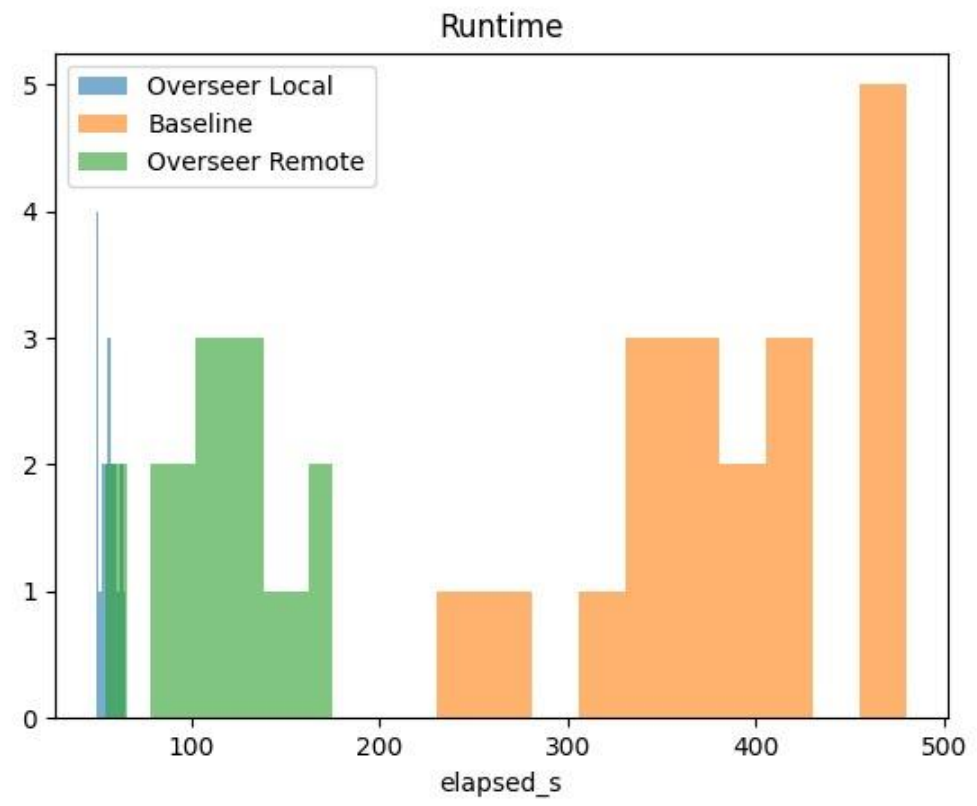

Figure 8**: Histogram of wall-clock runtimes for the Baseline and Overseer Local configurations on the dense matrix multiplication workload.**

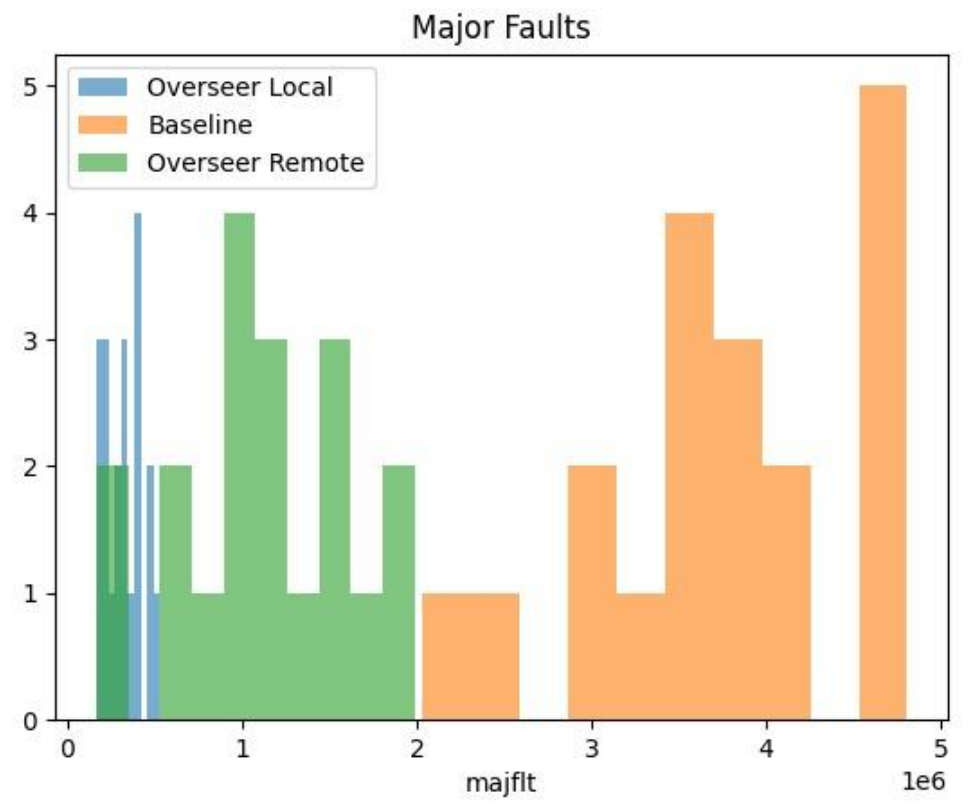

Figure 9: **Histogram of major page-fault counts for the Baseline and Overseer Local configurations on the Matrix Multiplication workload.**

Figures 8 and 9 show the runtime and major-fault distributions for Matmul. **Overseer Local** completes in roughly 50-60 seconds across all runs; **Overseer Remote** averages around 120-130 seconds and never exceeds 200 seconds, occasionally dipping into Local's range but not reliably so. **Baseline** never completes in under 200 seconds and clusters between roughly 250 and 500 seconds. Neither Overseer configuration overlaps with **Baseline** on runtime.

The major-fault picture is similarly separated: **Overseer Remote** stays below $2\times10^6$ faults per run while **Baseline** never drops below it. **Baseline** runs cluster between 3 and $4\times10^6$, with a further spike near $4.6\times10^6$ accounting for roughly a quarter of runs. **Overseer Local** sits lower still. The relationship between faults and runtime is near-linear, with each fault contributing enough time that even modest fault reductions produce visible runtime gains.

Both Overseer configurations sit well below **Baseline** despite Remote paying network and inference-latency costs Local does not. This is direct evidence that the page-fault latency budget absorbs an off-host inference hop without collapsing the win.

## 5.4 The Cost of Inference Latency

The gap between **Overseer Local** and **Overseer Remote** on Hilbert is one of the most informative results in this section. Both configurations share every design choice, every model weight, and every kernel-side mechanism; the primary difference is where **OverNet** executes and how predictions traverse to the Enforcer. The Local configuration delivers roughly twice the runtime reduction that the Remote configuration does, and roughly ten times the major-fault reduction. These results suggest two conclusions.

First, **Overseer's** achievable benefit is strongly influenced by prediction latency, rather than prediction quality alone. **OverNet's** predictions are accurate enough that, when they arrive promptly, they eliminate more than a quarter of the workload's major faults. When they arrive late, much of this potential benefit is lost in terms of fault avoidance level, though as noted above some of it is recovered as fault-servicing speedup.

Second, the disaggregated design's viability depends on the inference hardware being fast enough relative to the target's fault inter-arrival time. The Jetson Orin Nano configuration is viable in the sense that it delivers a clear speedup over **Baseline**, but a substantial fraction of the achievable benefit is left on the table by its inference latency. These results suggest that the disaggregated-inference approach can benefit directly from improvements in the attached inference hardware: a faster inference appliance could narrow the gap between Remote and Local without requiring changes to the **Collector** or Enforcer, and potentially without changes to the **OverNet** model architecture.

We consider this modularity an important property of the design. The mechanism side of the system, the mechanism-versus-policy separation described in §3.1, can accommodate improvements in inference hardware without requiring corresponding changes to the host-side data path. In contrast, on-die learned prefetchers face tighter model-capacity constraints imposed by the area and latency budgets established at design time.

# 6. Discussion and Conclusion

We presented **Overseer**, an end-to-end learned page-prefetching system that observes page faults via eBPF, predicts future page accesses with a transformer running on network-attached inference hardware, and materializes the predicted pages in the target's address space into the target's address space through a kernel-resident enforcer. The system splits cleanly along a mechanism-vs-policy boundary: the **Collector** and Enforcer form a workload-agnostic data path, while **OverNet**, trained on page-fault sequences and encoded through per-instance ordinal ranks over deltas, is the only workload-specialized element. Inference is disaggregated onto network-attached hardware, which removes prediction latency from the host's compute path and allowing the model to run on hardware optimized for ML inference.

Across two page-fault-intensive workloads, **Overseer** delivers substantial runtime improvements over the default Linux paging path. For Hilbert-curve traversal, local inference reduces mean runtime by approximately 35%, while remote inference on a Jetson Orin Nano achieves an approximately 18% reduction. For dense matrix multiplication, whose execution is more heavily dominated by page faults and whose fault stream is more predictable, local inference reduces mean runtime from approximately 450 s to 60 s, corresponding to a 7.5× speedup, while reducing major page faults by approximately an order of magnitude. The gap between the Local and Remote configurations on Hilbert further demonstrates the importance of inference-path latency: lower latency allows a larger fraction of accurate predictions to arrive early enough to avoid demand faults. More broadly, the results show that **Overseer's** disaggregated architecture can benefit from improvements in inference hardware without requiring corresponding changes to its host-side mechanism.

**Limitations. Overseer's** benefit depends on two properties of the target workload: the fraction of wall-clock time it spends on major faults, and the degree of structure its fault stream carries. Matmul is favorable along both dimensions and therefore achieves substantial improvement.; Hilbert is intermediate on both and sees a smaller but still substantial one. Workloads that fault rarely or whose faults are dominated by unpredictable access are likely to see little improvement and incur a small overhead for the observation path. The online vocabulary carries a warm-up cost during which the LookupTable is nearly empty and the resulting rank assignments carry limited information; on short-lived processes this warm-up may dominate. The Enforcer's kernel-resident UDP socket, while necessary for latency reasons, is a nontrivial expansion of the kernel's attack surface, and hardening it against malformed or malicious prediction packets is a real engineering concern rather than a solved problem. Finally, our evaluation covers two workloads on a single host and a single inference appliance; establishing broader generality will require evaluation across a more diverse workload suite and multiple hardware configurations.

**Future work.** Several directions follow naturally. The most immediate is characterizing the inference-latency frontier more thoroughly: at what point does a faster inference appliance stop translating into further runtime gains, and where does that saturation point sit relative to currently available accelerators. A second direction is online adaptation of OverNet itself, either through periodic fine-tuning against recently observed fault streams or through lighter-weight per-target adaptation layered on top of a shared base model. A third, and the one we consider most promising, is reimagining the **Collector** and Enforcer to target regimes beyond host-RAM paging on Linux. The mechanism-vs-policy split described in §3 was designed precisely to make this kind of retargeting possible: OverNet and its training pipeline are agnostic to where the fault stream comes from and where the predicted pages have materialized. Natural targets include accelerator-side page faults (an **Overseer** variant directed at GPU faults under Unified Virtual Memory would sit in a very different latency regime and interact with a very different memory hierarchy), and paging on tiered-memory or CXL-attached-memory hosts, where the cost of a miss is intermediate between a cache miss and a disk fault and the space of prefetch mechanisms is correspondingly richer.

The broader point **Overseer** makes is that the page-fault regime is a natural target for taking learned memory-policy components out of simulation and into deployment. The latency budget is generous enough to accommodate a real neural network with off-host inference, and the fault streams of realistic workloads are irregular enough that heuristic prefetchers leave meaningful headroom on the table. We hope **Overseer** serves as concrete existence proof and as a starting point for the systems questions that only become visible once a learned prefetcher is running.

## References

**[1]** Dongarra, Heroux & Luszczek, "High-performance conjugate-gradient benchmark," IJHPCA 30(1), 2015

[2] M. Hashemi, K. Swersky, J. A. Smith, G. Ayers, H. Litz, J. Chang, C. Kozyrakis, and P. Ranganathan. "Learning Memory Access Patterns." *Proceedings of the 35th International Conference on Machine Learning (ICML)*, 2018.

[3] Z. Shi, A. Jain, K. Swersky, M. Hashemi, P. Ranganathan, and C. Lin. "A Hierarchical Neural Model of Data Prefetching." *Proceedings of the 26th ACM International Conference on Architectural Support for Programming Languages and Operating Systems (ASPLOS)*, 2021.

[4] P. Zhang, A. Srivastava, A. V. Nori, R. Kannan, and V. K. Prasanna. "Fine-Grained Address Segmentation for Attention-Based Variable-Degree Prefetching." *Proceedings of the 19th ACM International Conference on Computing Frontiers (CF)*, 2022.

[5] Q. Duong, A. Jain, and C. Lin. "A New Formulation of Neural Data Prefetching." *Proceedings of the 51st Annual International Symposium on Computer Architecture (ISCA)*, 2024.

[6] P. Michaud. "Best-Offset Hardware Prefetching." *Proceedings of the 22nd IEEE International Symposium on High Performance Computer Architecture (HPCA)*, 2016.

[7] A. Jain and C. Lin. "Linearizing Irregular Memory Accesses for Improved Correlated Prefetching." *Proceedings of the 46th Annual IEEE/ACM International Symposium on Microarchitecture (MICRO)*, 2013.

[8] M. Bakhshalipour, P. Lotfi-Kamran, and H. Sarbazi-Azad. "Domino Temporal Data Prefetcher." *Proceedings of the 24th IEEE International Symposium on High Performance Computer Architecture (HPCA)*, 2018.

[9] X. Long, X. Gong, B. Zhang, and H. Zhou. "Deep Learning based Data Prefetching in CPU-GPU Unified Virtual Memory." *Journal of Parallel and Distributed Computing*, vol. 174, pp. 19-31, 2023.

[10] Y. Wang, Y. Meng, J. Wang, and C. Yang. "LSTM-CRP: Algorithm-Hardware Co-Design and Implementation of Cache Replacement Policy Using Long Short-Term Memory." *Big Data and Cognitive Computing*, vol. 8, no. 10, p. 140, 2024.

[11] A. Vaswani, N. Shazeer, N. Parmar, J. Uszkoreit, L. Jones, A. N. Gomez, Ł. Kaiser, and I. Polosukhin. "Attention Is All You Need." Advances in Neural Information Processing Systems, vol. 30, 2017.